\documentclass[conference]{IEEEtran}
\IEEEoverridecommandlockouts
\usepackage{cite}
\usepackage{amsmath,amssymb,amsfonts}
\usepackage{algorithmic}
\usepackage{graphicx}
\usepackage{textcomp}
\usepackage{xcolor}
\usepackage[pangram]{blindtext}

\def\BibTeX{{\rm B\kern-.05em{\sc i\kern-.025em b}\kern-.08em
    T\kern-.1667em\lower.7ex\hbox{E}\kern-.125emX}}

\usepackage[utf8x]{inputenc}

\makeatletter
 \let\old@ps@headings\ps@headings
 \let\old@ps@IEEEtitlepagestyle\ps@IEEEtitlepagestyle
 \def\confheader#1{%
 \def\ps@headings{%
 \old@ps@headings%
 \def\@oddhead{\strut\hfill#1\hfill\strut}%
 \def\@evenhead{\strut\hfill#1\hfill\strut}%
 }%
 \def\ps@IEEEtitlepagestyle{%
 \old@ps@IEEEtitlepagestyle%
 \def\@oddhead{\strut\hfill#1\hfill\strut}%
 \def\@evenhead{\strut\hfill#1\hfill\strut}%
 }%
 \ps@headings%
 }
 \makeatother
 
\confheader{- Preprint submitted to the 34\textsuperscript{th} International Conference on Software Maintenance and Evolution (ICSME'18) -}

\begin{document}

\title{Aligning Technical Debt Prioritization with Business Objectives: A Multiple-Case Study
}

\author{
    \IEEEauthorblockN{Rodrigo Rebouças de Almeida\IEEEauthorrefmark{1}, Uir\'{a} Kulesza\IEEEauthorrefmark{1}, Christoph Treude\IEEEauthorrefmark{2},\\ D'angellys Cavalcanti Feitosa\IEEEauthorrefmark{3}, Aliandro Higino Guedes Lima\IEEEauthorrefmark{4}}
    \thanks{\IEEEauthorrefmark{1}Rodrigo Rebouças de Almeida is also an associate professor at the Federal University of Parai­ba - UFPB, Rio Tinto, Brazil.}
	\thanks{This work was partially supported by CAPES, PPgSC-UFRN and National Institute of Science and Technology for Software Engineering (INES 2.0) - CNPq grant 465614/2014-0.}
    \IEEEauthorblockA{\IEEEauthorrefmark{1}PPgSc - Federal University of Rio Grande do Norte - UFRN, Natal, Brazil}
        \IEEEauthorblockA{\IEEEauthorrefmark{2}University of Adelaide, Adelaide, Australia}
        \IEEEauthorblockA{\IEEEauthorrefmark{3}Conductor, São Paulo, Brazil} 
        \IEEEauthorblockA{\IEEEauthorrefmark{4}Dataprev, Brasilia, Brazil \\
rodrigor@dcx.ufpb.br, uira@dimap.ufrn.br, christoph.treude@adelaide.edu.au, \\ dangellys.feitosa@conductor.com.br, aliandro.lima@dataprev.gov.br} 
}

\maketitle

\begin{abstract}
Technical debt (TD) is a metaphor to describe the trade-off between short-term workarounds and long-term goals in software development. Despite being widely used to explain technical issues in business terms, industry and academia still lack a proper way to manage technical debt while explicitly considering business priorities.
In this paper, we report on a multiple-case study of how two big software development companies handle technical debt items, and we show how taking the business perspective into account can improve the decision making for the prioritization of technical debt. We also propose a first step toward an approach that uses business process management (BPM) to manage technical debt.
We interviewed a set of IT business stakeholders, and we collected and analyzed different sets of technical debt items, comparing how these items would be prioritized using a purely technical versus a business-oriented approach.
We found that the use of business process management to support technical debt management makes the technical debt prioritization decision process more aligned with business expectations. We also found evidence that the business process management approach can help technical debt management achieve business objectives.\end{abstract}

\begin{IEEEkeywords}
Technical Debt Management, Technical Debt, Business Process Management, Technical Debt Prioritization
\end{IEEEkeywords}

\section{Introduction}

The technical debt (TD) metaphor, coined by Cunningham~\cite{Cunningham:1992}, has been used to describe the trade-off between short-term benefits gained by delaying certain development activities and the costs of implementing these activities in the future. Although the metaphor has been used to facilitate the communication between developers and business stakeholders, the state of the art of technical debt management is still in need of a more appropriate treatment of how technical debt affects the business, and vice versa~\cite{seaman:guo:2011,Siebra:2012,Klinger:2011,LI:2015}. 

Since one of the primary sources of technical debt are business forces, such as time to market and customer satisfaction~\cite{FERNANDEZSANCHEZ201722}, a new perspective on technical debt management, built upon business values and priorities, is needed.

The report from Dagstuhl Seminar 16162~\cite{avgeriou_et_al:2016}, which involved 33 researchers, practitioners and tool vendors from academia and industry, presents a research roadmap for technical debt. Its authors argue that ``business value is central to delivering effective mechanisms for managing TD in practice''. They also state that ``demonstrating the benefits of considering TD in management decisions is a key area for TD researchers''.

Technical debt management has received attention from both industry and the research community. Several approaches have been proposed to deal with decision making in technical debt management. Guo and Seaman~\cite{seaman:guo:2011} propose a portfolio approach, and Kruchten at al.~\cite{Kruchten:2012} provide a technical debt landscape and propose four types of possible improvements, classified as positive, negative, visible, and invisible, to organize technical debt items in backlogs.

 Guo and Seaman~\cite{Guo:2011, seaman:guo:2011} propose a technical debt management framework centered around a technical debt list. The technical debt list contains a set of technical debt items, which represent an instance of technical debt. The framework also considers the principal (the cost of fully eliminating the debt), the interest (the additional cost of postponing the task), and the interest probability.

In this work, we collected and analyzed a set of 188 technical debt items from two large software development companies and conducted a focus group and interviews with IT and business stakeholders to understand how taking the business perspective into account can improve the decision making related to technical debt prioritization. As a result of our case study, we also propose an extension of Seaman and Guo's framework~\cite{seaman2011} by using business process management (BPM)~\cite{bpmdumas} to improve the understanding of how critical or urgent a technical debt item is.
Our results show evidence that taking business priorities into account can change decisions related to technical debt prioritization. To the best of our knowledge, this is the first work which uses business process management to support decision making in technical debt management.

The main aim of our study is to understand how taking the business perspective into account can affect the prioritization of technical debt items. We set out to answer the following research questions:
\begin{itemize}
  \item \textbf{\textit{RQ 1. How can the business perspective influence the prioritization of technical debt?}} To answer this research question, we interviewed IT and business stakeholders from two large software development companies, and we collected and analyzed a set of 188 technical debt items from different systems, comparing how these items would be prioritized using a purely technical approach versus a business-oriented approach.
  \item \textbf{\textit{RQ 2. Does the business perspective captured through business process management facilitate the prioritization of technical debt?}} We explore how business process management (BPM) can contribute to making technical debt prioritization more aligned with the business objectives. Fourteen different business processes in two cases were identified, and some of them were modeled. These business process models were used to analyze how the information about two business metrics (criticality and urgency) contributes to the technical debt prioritization.
  \end{itemize}

Our results show that using business process management to capture the business perspective facilitates the prioritization of technical debt in order to address business expectations. It also helps to improve the argumentation from the technical side to convince business stakeholders to prioritize what was previously considered pure-technical problems.

\section{Case Study}

The objective of this study was to investigate whether business process management is a viable tool to support technical debt decision making. To answer our research questions, we conducted a multiple-case study~\cite{case.study.yin} with two software development companies.

\subsection{Theoretical basis}

\begin{figure}[ht]
\begin{center}
\includegraphics[width=0.9\columnwidth]{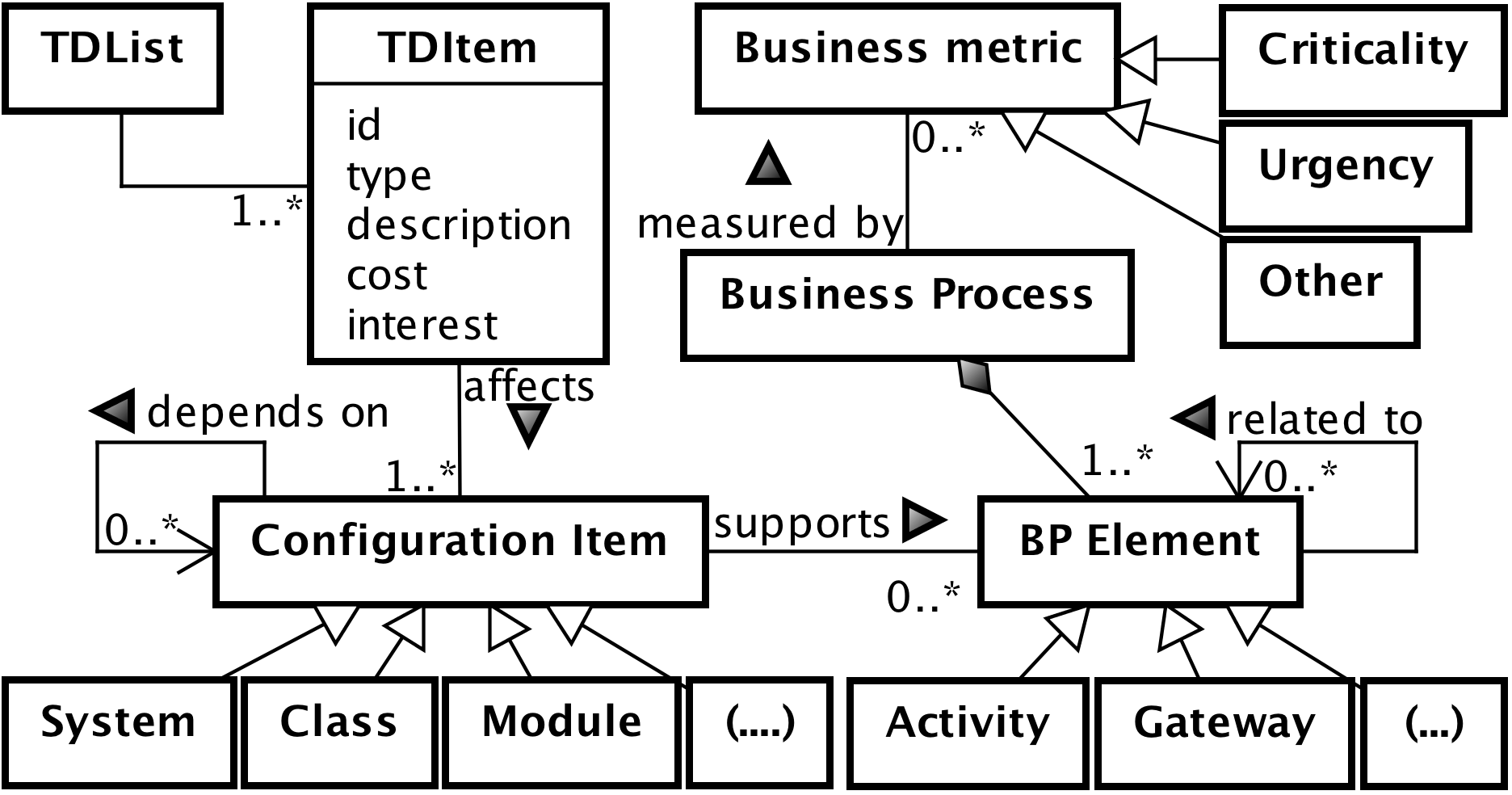}
\caption{Relationships between the technical debt list and the business process}
\label{fig_uml}
\end{center}
\end{figure}

In this work, we collect the business perspective of technical debt items through their relationship with business processes.
Figure~\ref{fig_uml} shows our conceptual model of how technical debt items and business processes are related to each other. The model shows that a technical debt list ``TDList'' is related to one or more technical debt items ``TDItems'' which affect one or more ``Configuration Items''. A configuration item can be any technical artifact or system or service which is directly or indirectly affected by technical debt. For example, a test debt item can affect a Java class which can affect a system module, which then can affect an IT service, which can support a business activity, and finally a business process. All items, from the Java class to the IT service, are instances of configuration items.
A configuration item can support different business process elements ``BP Elements''. Business processes are what companies do to deliver value to customers. For example, a ``sales'' process in an e-commerce company is the set of activities, decisions, and events that must happen to allow the customer to buy products~\cite{bpmdumas}.
A BP Element can have its priority and criticality evaluated in business terms. BP Elements compose the ``Business Process'', which also has its overall priority and criticality. This model extends the conceptual model presented by Rios et al.~\cite{Rios2018} by adding the business process perspective.

\subsection{Case Study Design}

The objective of our case study was to gather data about current technical debt from software development companies, to understand how the systems and services affected by the debt support business processes, and to understand the business processes priorities and if these priorities would affect technical debt prioritization decision making. For the case study, we followed the steps outlined by Runeson, H\"{o}st~\cite{case.study.runeson} and Yin~\cite{case.study.yin}. First, we planned the case study, designed it, prepared the protocol, collected data and finally, analyzed the results. The following subsections will detail each step. 

\subsubsection{Requirements for the case study}

We had the following \textbf{requirements for teams to participate in the case study}:

\begin{itemize}
	\item \textbf{Availability:} the team must be available to participate in the research, to give access to data and allow the execution of activities such as interviews, focus groups, and observations. In addition, the company must provide access to pure-business, management, and technical stakeholders.
	\item \textbf{Suffer from technical debt and maintain a list of debt items}: (this was the easiest requirement to meet) the team must understand what technical debt is and maintain a list of technical debt items to be handled by the team. This requirement was essential to avoid research bias: If the teams had not had an existing list of technical debt items, creating such a list for the purpose of this research could have interfered with the perception of priorities.
	\item \textbf{Be exposed to direct business pressures:} the team must be affected by business stakeholders in their day-to-day work. This requirement excludes teams who work on systems which do not have direct business impact, e.g., teams working on infrastructure. 

\end{itemize}

\subsubsection{Selected cases}

We selected two teams from two companies that were part of a set of eighteen industry partners which collaborate with our research group. In this paper, we will refer to the companies as ``Company A'' and ``Company B'', to their teams as ``Team A'' and ``Team B'', and the cases as ``Case A'' and ``Case B''.

Both companies are typical software development companies which develop systems for third-party customers. Company A is part of the government, and Company B is private and provides solutions for credit card processing (private label and co-branded). 
Company A has more than 600 developers and is responsible for the development and service support of more than 200 different products for different government customers. They handle a country-scale data set. In their case, a product is a set of IT solutions in the scope of a business contract. Their products comprise information systems, mobile systems, data processing, and business intelligence solutions.

Company B has 450 employees, more than 300 different projects, and around 95 different clients. The company is focused on solutions for credit card processing. It processes a mean of 2 million transactions per day, accounting for around 130 million dollars per month.

After selecting the companies, we selected teams suitable for our study. Both selected teams develop large-scale software systems and use a commonly-used technology stack and architecture. Team A develops transaction-intensive information systems and does large-scale data processing. Team B works with large-scale transaction processing systems, on private-label credit card processing.

Team A is composed of 22 professionals with roles such as service support, software developer, software architect, technical leader, system analyst, service manager, and account manager. Team B is composed of 8 professionals, with an agile flavor: software developer, technical leader, test analyst, and product owner.

Both teams use a SCRUM-like development process, they develop systems with high business impact, and are directly affected by business pressures.

\textbf{Regarding technical debt management}, the individuals of both teams understand the term and routinely handle cases of technical debt. Team A developed an internal tool to track and prioritize technical debt items. They also established best practice guidelines for developers, and periodically, there is a team of technical leaders who analyze the code produced by their teams looking for technical debt items. 

Team B tracks technical debt items using the task management tool Trello. They use this tool in technical meetings, primarily when the technical debt items are responsible for incidents or are delaying the implementation of features. 

\subsection{Data collection and analysis protocol}

\begin{figure}[ht]
\begin{center}
\includegraphics[width=0.9\columnwidth]{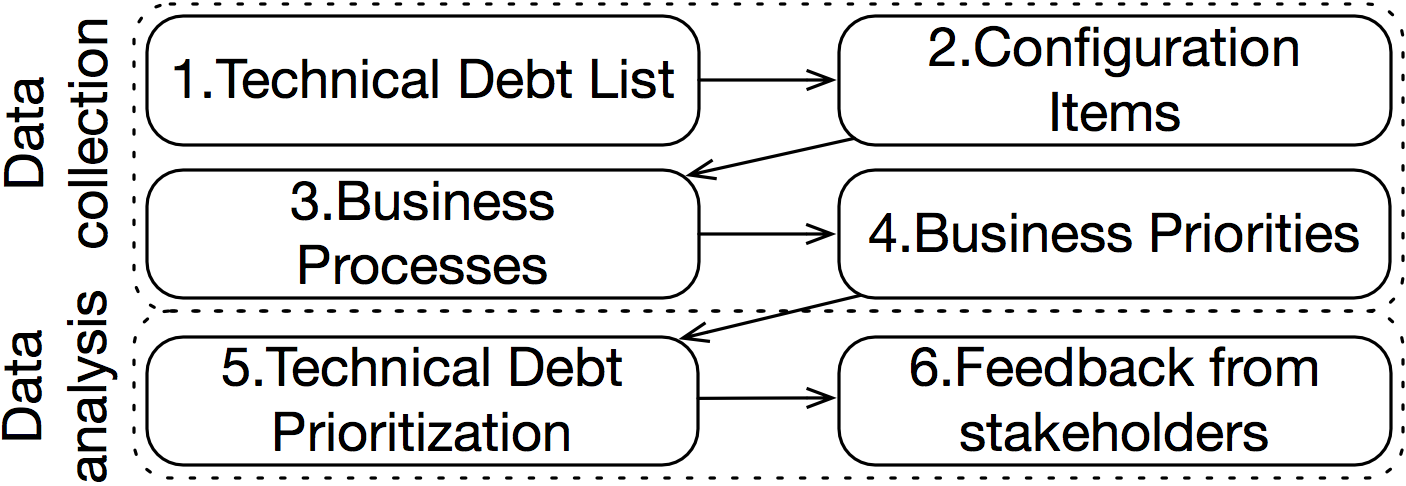}
\caption{Data collection and analysis main steps}
\label{fig_protocol}
\end{center}
\end{figure}

In this section, we present the data collection and analysis protocol applied in both cases. Figure~\ref{fig_protocol} presents the six steps of data collection and analysis. We first collected a list of technical debt items from both teams and worked with the team members to identify the debt items' priorities from a technical point of view as well as the configuration items affected by the debt items. We then worked with business stakeholders from both teams to identify and model the business processes affected by these configuration items and we identified the priorities of the corresponding activities. To prioritize the technical debt items from the point of view of business objectives, we then mapped the prioritized business processes to the list of technical debt items, using the configuration items. This enabled us to compare the technical debt prioritization from both viewpoints: the business perspective and the technical perspective. Finally, we discussed results with stakeholders from both sides. We describe each step in detail in the following.

\subsubsection{Technical debt list}

\begin{table}[ht]
\begin{center}
\includegraphics[width=0.7\columnwidth]{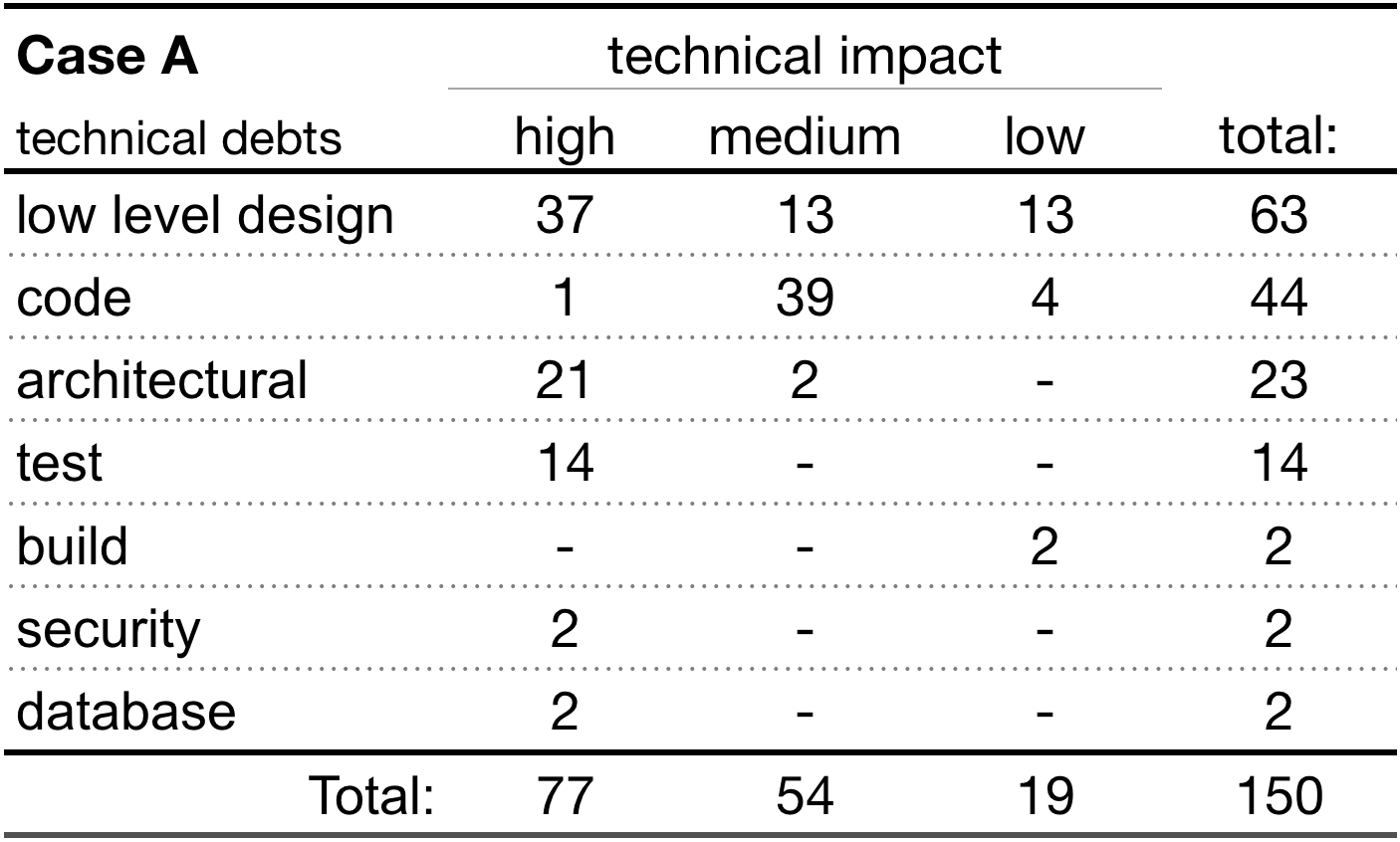}
\caption{Technical debt categories and impact - Case A}
\label{table_tdtype.teamA}
\end{center}
\end{table}

\begin{table}[ht]
\begin{center}
\includegraphics[width=0.7\columnwidth]{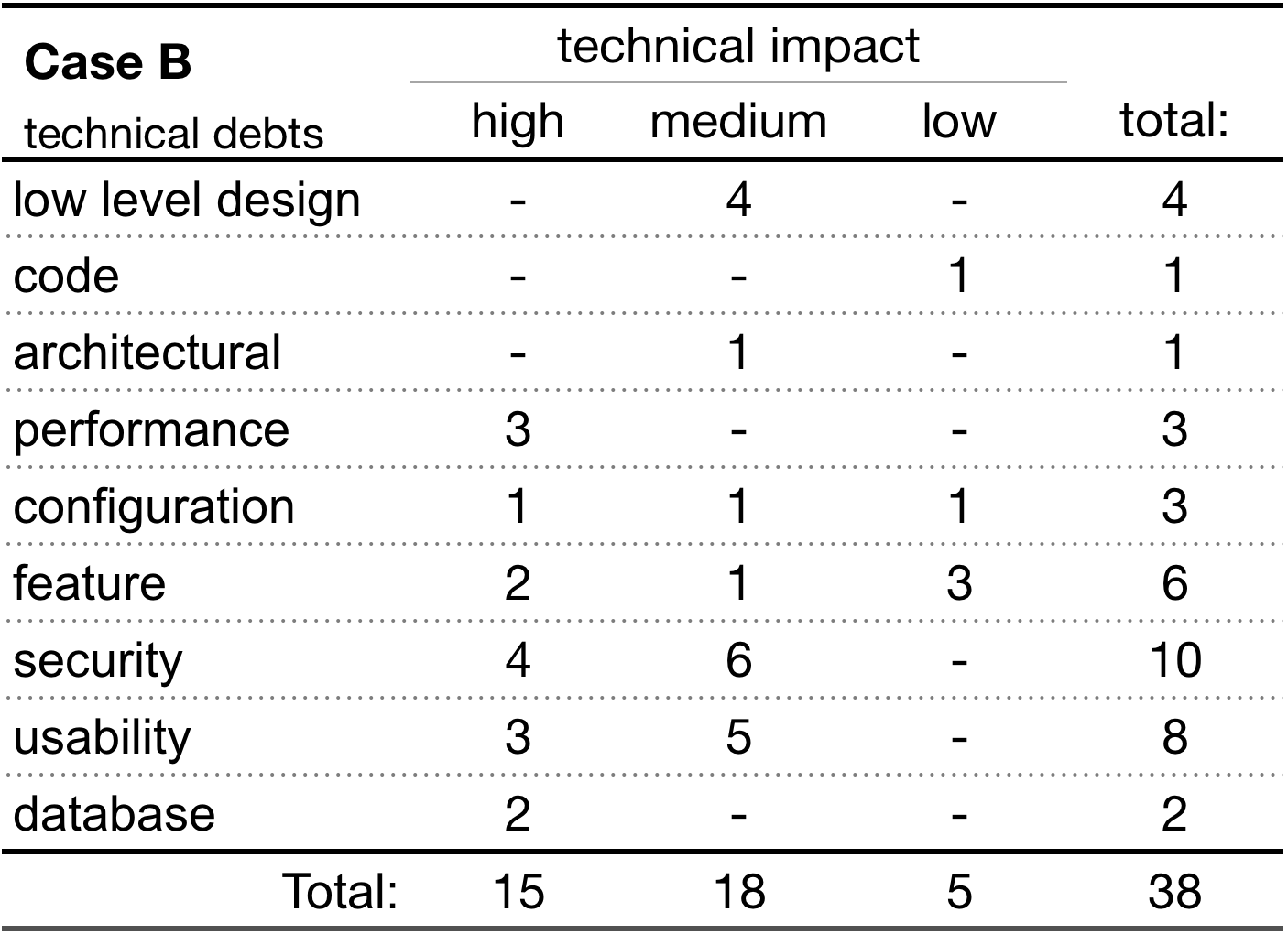}
\caption{Technical debt categories and impact - Case B}
\label{table_tdtype.teamB}
\end{center}
\end{table}

To compare the differences between a technical prioritization and a business-oriented one, we collected a set of technical debt items prioritized according to their impact. \textbf{The impact of technical debt is the amount of the consequences of not paying the debt~\cite{seaman:guo:2011}}. In other words: what happens if the debt is not paid? To gather this data, we collected a high/medium/low evaluation from technical stakeholders (technical leaders, experienced developers or software architects, for example). We describe the details of this data collection for each case in the following paragraphs.

For Team A, we obtained access to a set of 150 technical debt items with their full descriptions, annotations, and classifications. Table~\ref{table_tdtype.teamA} presents a summary of this data. It shows how the technical debt items were ranked regarding their ``impact''. Note that this categorization had been made before our study by the technical leaders of Team A who registered the items. We found 19 items with low impact, 54 items with medium impact, and 77 items with high impact. The full technical debt list is available in the companion data \cite{companionData}.

Team B did not have an explicit list of technical debt items, which required us to analyze a set of 249 user stories and 173 issues from their task management system to select them. After two meetings with a technical leader, we selected 25 technical debt items for the case study.
Since the items had not been previously prioritized, we conducted a focus group with seven technical team members (developers, technical leader, and system analyst) with the objective to prioritize the impact of the selected technical debt items from a technical perspective. Each team member classified the technical impact of all items individually at first and subsequently collaboratively discussed any divergences and agreed on the final prioritization.

Table~\ref{table_tdtype.teamB} presents a summary of the 38 technical debt items of nine different types. 5 were classified as low impact, 18 as medium and 15 as high.

\subsubsection{Configuration items} After obtaining the list of technical debt items, we scheduled meetings with technical leaders and senior developers on the two cases to identify which configuration items were affected by each technical debt item.

On Team A, the information about the affected services was identified by the technical leader during code review, for each technical debt item. For example, a particular security debt item had this comment: ``Occurrences: PayrollServices.replace, PayrollServices.updatePayrollWithTotalValue, (...)''. The comment refers to a Java class which implements a JEE service. We analyzed all technical debt items, identified from the comments all occurrences of each technical debt item and asked the software architect to identify which systems and services were being affected by these occurrences. We conducted three meetings to cover all 150 technical debt items. In the end, they were mapped to five information systems and three batch jobs. A technical leader, with long-time experience on the project, also verified the mapping, adjusting a few mappings and ultimately agreeing on the final result. The companion data  \cite{companionData} has information about the identified configuration items in Case A.

For Team B, we identified the configuration items affected by each technical debt item by reading the item descriptions and verifying our understanding with a technical leader. Note that the configuration items in Case B were at a higher level (i.e., systems) of abstraction compared to Case A (classes and modules).

\subsubsection{Business Processes} 

To understand how the previously identified configuration items support the business, we identified and modeled the business processes supported by the configuration items affected by each technical debt. 

In both cases of our study, the team had only used the artifacts related to the description and analysis of business processes in the initial phases of their project (scope definition and high-level analysis). Neither case had structured documentation of the business processes supported by the systems and services, i.e., we had to model the business processes in collaboration with business stakeholders and senior tech leaders. 

The business process modeling was done in both teams in two steps. First, we interviewed a system analyst to obtain information about the business processes and modeled them according to Silver~\cite{bpmnstyle} and Dumas et al.~\cite{bpmdumas}. The processes were modeled using BPMN 2.0~\cite{bpmn2.0}. A project manager validated them and gave input for adjustments. For each case, the output was a detailed business process model with activities and decisions about internal procedures.

In Case A, technical debt items affect the systems that support a large business process that was detailed in three subprocesses, see Table~\ref{table_c1.bps}. In Case B, 13 processes were identified and one was detailed, see Table~\ref{table_c2.bps}.

In Team A, the model was validated by the account manager, in a semi-structured meeting where a researcher presented the business process model and guided the discussion for each business activity. The account manager could make comments and present her concerns about the current modeling.

The process modeled for Team A has three main subprocesses: ``Request for Payment'', ``Customer Service'', and ``Payment'' (Table  \ref{table_c1.bps}). In the process, citizens request a financial benefit (``Request for Payment''), then go to a service center and provide documentation to a ``pre-qualification'' subprocess, which analyses if they can receive the requested payment. After that, if they are qualified, they are forwarded to be processed in ``Service A'' (``Forward citizen to Service A'').

The ``Request for Service'' subprocess is responsible for the processing of an average of more than 1.2 million requests per month. The requests are made using a web application on the Internet. The ``Customer Service'' subprocess is responsible for an average of 30,000 requests per day, in around 1,500 service centers across the country. The ``Payment'' subprocess is responsible for the processing of a mean of 700,000 payment orders and handles around US\$ 270,000 per week.

In Team B, the models were validated by a senior analyst, who had in-depth knowledge about the business side of the project. The validation meeting was also a semi-structured interview, where each process and activity was revised. In this case, after the meeting, we identified 13 business processes directly affected by the five systems (i.e., configuration items). Table~\ref{table_c2.bps} enlists the set of 13 business processes and the 8 activities from the ``Invoice payment and scheduling'' process. 

\begin{figure*}[ht]
\begin{center}
\centerline{\includegraphics[width=0.9\textwidth]{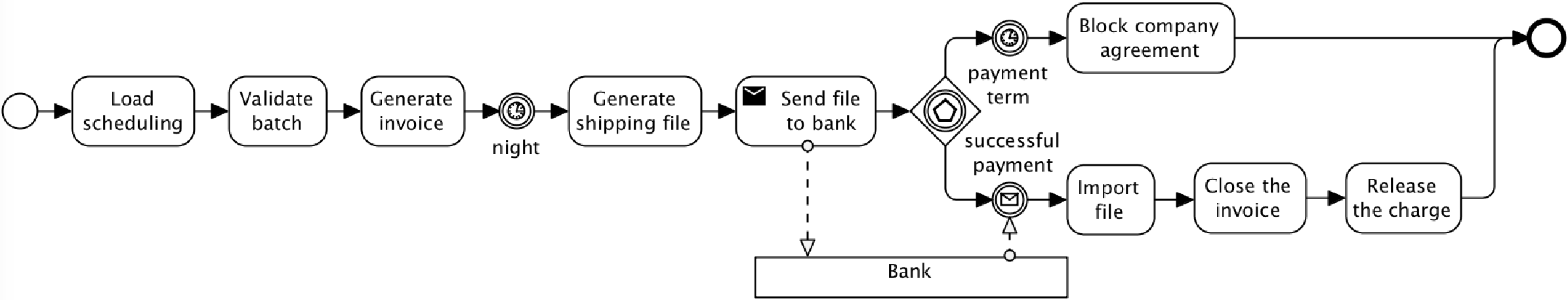}}
\caption{Business process model example: ``Invoice payment and scheduling'' - Case B}
\label{fig_bpmn}
\end{center}
\end{figure*}

In this second case, the majority of the business processes could be evaluated as a black box, i.e., without details about activities, events, and decisions. This was possible when a single system or module automated all activities of the process, i.e., if the system or model is affected, the whole process is affected. ``5. Card sale'' is an example of a highly critical and urgent business process supported by a single system. This was not the case for ``Invoice payment and scheduling'', where different activities had different urgency and criticality. Figure~\ref{fig_bpmn} shows the activities and decisions from the moment the company schedules a set of payments to be credited to their employees to the moment that the credit is charged to their credit cards.

\subsubsection{Business Priorities}

\begin{table}[ht]
\begin{center}
\includegraphics[width=0.8\columnwidth]{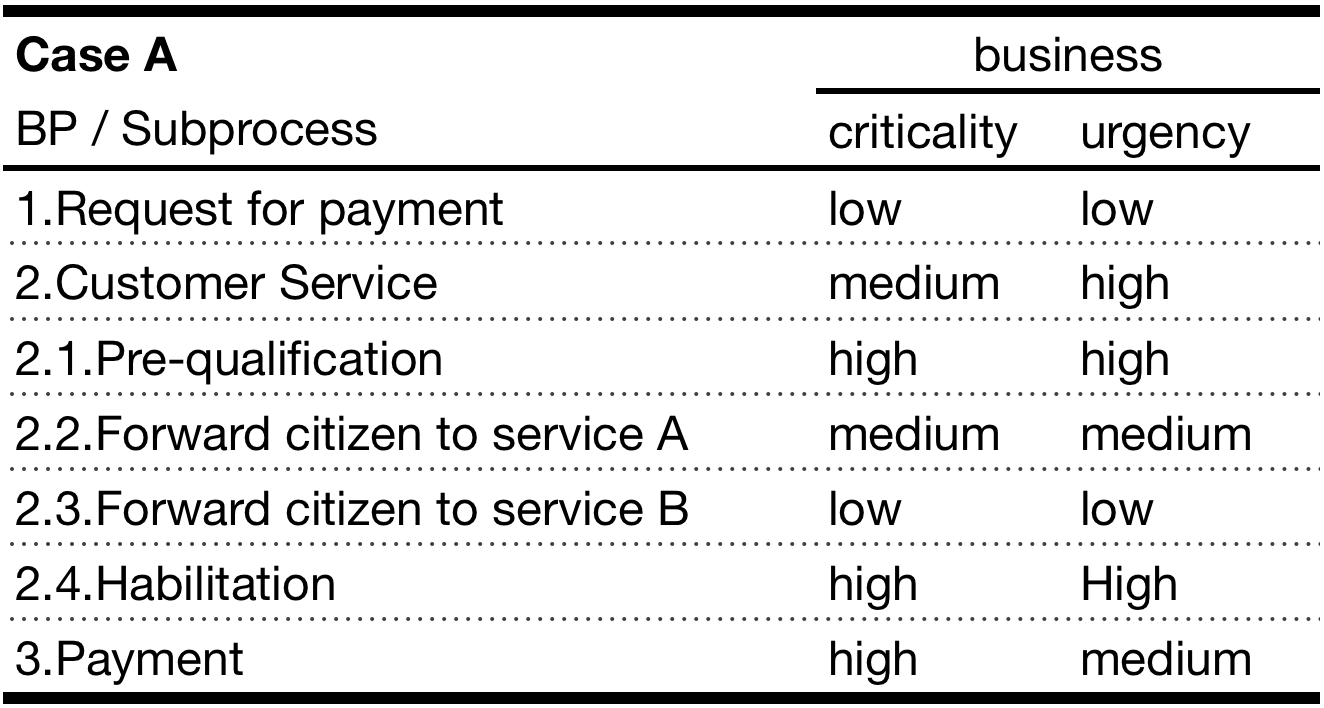}
\caption{Business processes and their urgency and criticality - Case A}
\label{table_c1.bps}
\end{center}
\end{table}

\begin{table}[ht]
\begin{center}
\includegraphics[width=0.8\columnwidth]{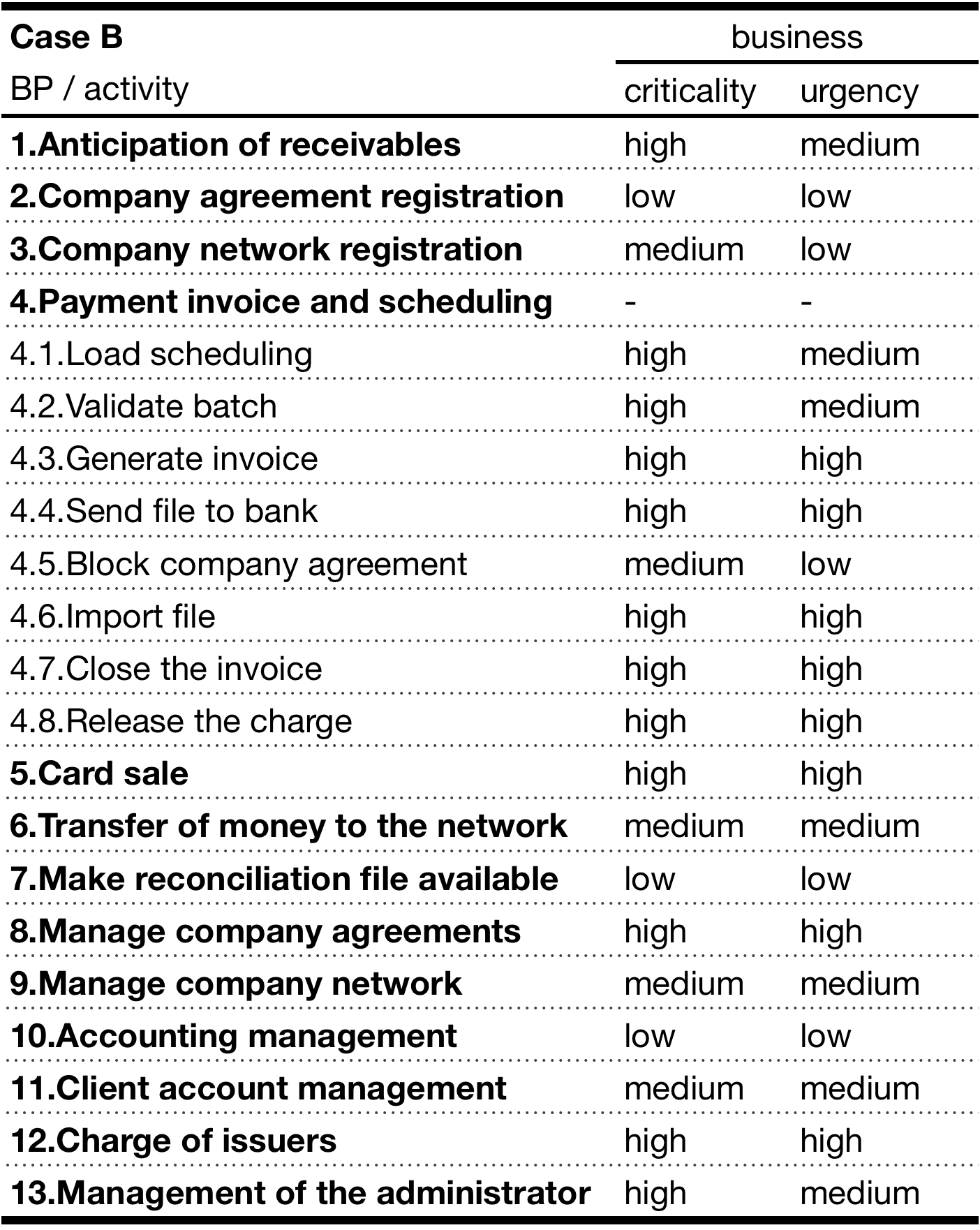}
\caption{Business processes and their urgency and criticality - Case B}

\label{table_c2.bps}
\end{center}
\end{table}

The next step was to prioritize the business process activities from a business perspective. During semi-structured interviews, we asked business stakeholders to provide their perception of criticality and urgency of the business processes. Sometimes they classified the whole business process (in the case where all activities within a process had the same classification), and sometimes they classified specific activities within a process (when different activities had different classifications).

For Case A, we asked the account manager -- a business stakeholder -- to analyze the modeled business process and evaluate each subprocess. The business criticality was evaluated considering the business value of each subprocess for the citizens while the business urgency was evaluated considering how fast a problem must be solved in order to reduce impact on citizens. The account manager also evaluated the urgency and criticality of the subprocesses of the ``Customer Service'' process. The final business prioritization is shown in Table \ref{table_c1.bps}.

Different from Case A, where the business priorities were a measure of how the process affects citizens, in Case B (see Table \ref{table_c2.bps}), the business process priorities were evaluated considering their impact on revenue and the relationship with business partners. For example, the ``Card sale'' business process, which enables the customer buying activity, was evaluated as highly critical and highly urgent. ``If this process is affected by some problem, the customer can't use their card'', argues the business analyst. The system which supports this process also has an availability service level agreement (SLA) of 99.98\%. 
The ``4.Payment invoice and scheduling'' business process has a set of activities which have different criticalities and urgencies. Many of its activities with medium or low urgencies are due to the implementation of automated redundancy or there is a way to run actions manually, e.g., ``4.5 Block company agreement''.

\subsubsection{Technical Debt Prioritization}

To prioritize the technical debt items from the point of view of business objectives, we created a new technical debt prioritization, considering the business criticality and urgency. The same procedure was executed in both cases: we mapped the prioritized business processes to the list of technical debt items, using the configuration items. Then we compared the technical debt prioritization from both viewpoints: the business perspective and the technical perspective.

\subsubsection{Feedback from Stakeholders} \label{feedback}

After prioritizing the technical debt items using the business perspective, we ran a set of semi-structured meetings with pure-business and technical stakeholders, to discuss the results. All conversations in these meetings and interviews were recorded and summarized into higher-level themes by the first author as part of a qualitative analysis. The findings described in the Results section capture these themes.

The meetings had the following structure:
\begin{itemize}
	\item Show the list of technical debt items and the evaluation of their technical impact. We selected two examples to present in detail. Then we asked if participants understood them and if they had any question about the examples. 
	\item We then presented the technical debt items ordered by their technical impact. 
	\item Next, we showed the list of business processes affected by the technical debt items in the scope of the case study. We asked participants to review the business processes and asked if there is any concern regarding their criticality and urgency ratings.
	\item Lastly, we presented the prioritization considering the business perspective and compared it with the prioritization using the technical perspective. We asked if participants had any questions about the new prioritization and we asked if the presented perspective would be useful when handling technical debt items. Finally, we asked them for comments.
\end{itemize}

In Team A, we ran this meeting three times, one with the account manager (a pure-business stakeholder), one with a software architect, and one with a project manager.

In Team B, we ran this meeting a total of five times: first with the business stakeholder who helped with the business process description and second with a senior developer. We then followed the same meeting structure with 3 additional product owners of 3 different projects from the same company. Since they were from the same company and even though they could not evaluate the accuracy of the technical and business evaluation, they understood the problem and the proposed solution and could evaluate the prioritization using a business perspective. 

To evaluate how business and technical stakeholders would use the results from the case study to decide which technical debt items should be selected and how these items should be prioritized in a conflict scenario between business and technical interests, we conducted an additional focus group in Team B. One pure-business stakeholder and one senior developer participated in this focus group. Both stakeholders had more than ten years of experience and had worked with the business model for more than four years. 

The focus group was divided into two rounds. In the first round, the participants had access to the 38 technical debt items (each technical debt item had information about its technical impact and its business criticality and urgency); and both participants were asked to select and prioritize ten debt items to be the scope of development in the following development sprints.

In the second round, they were asked to consider their 10 (a total of 20) debt items and negotiate to choose which 10 would be part of the final selection. After the first round, only one technical debt item was selected by both the business and technical participant. After their negotiation, they identified the final selection and prioritization as shown in Table \ref{table_finalPriority}. Note that in Case B due to the nature of their debt items, we treated technical debt items affecting multiple business processes as separate items. For example, a highly generic debt item such as ``We need a security solution'' was broken up into the need for a security solution for System A, for System B, etc. The list of the selected technical debt items is available in the companion data~\cite{companionData}.

\section{Results}

In this section, we present and discuss the answers to our research questions. 

\begin{table}[ht]
\begin{center}
\includegraphics[width=0.7\columnwidth]{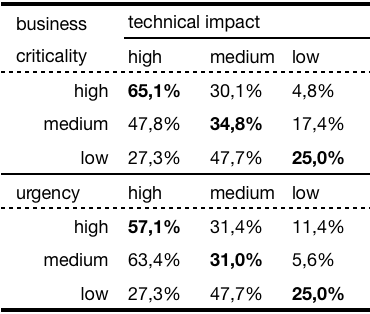}
\caption{Technical impact versus business impact - Case A}
\label{table_c1_impactVsBusiness}
\end{center}
\end{table}

\begin{table}[ht]
\begin{center}
\includegraphics[width=0.7\columnwidth]{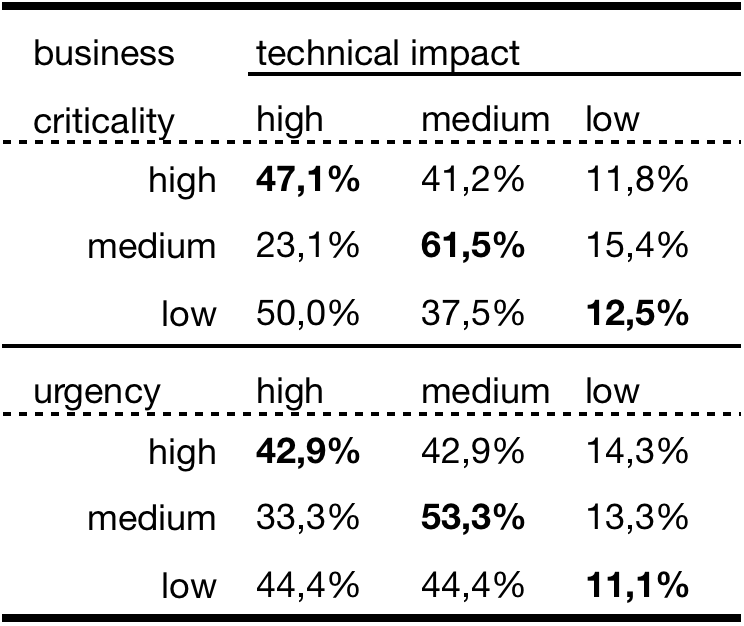}
\caption{Technical impact versus business impact - Case B}
\label{table_c2_impactVsBusiness}
\end{center}
\end{table}

\subsection{RQ 1. How can the business perspective influence the prioritization of technical debt?}

With the business prioritization for each process, subprocess and activities in hand and all technical debt items linked to their corresponding business entities (processes, activities, and so on), we step forward to the new technical debt prioritization, considering the business perspective.

Table~\ref{table_itemExample} shows a subset of technical debt items from Case A, each with technical impact, business criticality, and urgency rankings. The table shows the items ordered by their criticality, with higher criticality first. Note the differences and conflicts between technical and business perspectives.

Tables~\ref{table_c1_impactVsBusiness} (Case A) and~\ref{table_c2_impactVsBusiness} (Case B) show the percentage of technical debt item priorities which matched the business expectation. They show how misaligned this decision would be with business objectives if the team would prioritize the technical debt considering only a technical perspective. 

In Case A (Table~\ref{table_c1_impactVsBusiness}), regarding business criticality, 65\% of the technical debt items classified as high priority matched the business expectation. The same applies to 34.8\% of the medium priority items and 25\% of the low priority items. In total, the technical prioritization matched only 48.7\% of the criticality prioritization and only 35\% matched the urgency expectation. \textbf{This result provides evidence on how different a purely technical prioritization could turn out if it had been conducted from a business perspective.}

In Case A, for example, 34.8\% (30.1\% medium + 4.8\% low) of the technical debt items which affect highly critical business processes would not be classified as high priority. Instead, 27.3\% of the high impact technical debt items, which affect non-critical business processes, would be prioritized. If we consider the urgency to solve problems on business processes, also in Case A, the situation would be worse, since 42.8\% (31.4\% medium + 11.4\% low) of the technical debt items which affect business processes with high urgency would not be prioritized.

In Case B, (Table~\ref{table_c2_impactVsBusiness}), we can see that 87.5\% of the debt items ranked as medium and high affect business processes with low criticality, while 52\% of the debt items that affect highly critical business processes are not ranked as having a high technical impact. 

It is clear that we would not expect a complete correspondence between the technical and the business perspectives, since the technical aspects which guide the prioritization are different from the business aspects which guide business prioritization. However, the results show that a business-driven prioritization, through the business process perspective, can be useful to support the prioritization of technical debt.

When we showed these results to both business and IT stakeholders in Case A, they understood that something was missing in what was being considered in their prioritizations. The result does not mean that we should consider a purely business-focused perspective when prioritizing a technical debt item, nor a purely technical one. We should consider the trade-offs of each situation to find a balance to enable efficient decision making.

In Case B, the team members mentioned an opportunity to expand the metrics from the high/medium/low ranking to a financial metric in the future. They also saw opportunities to help with scope negotiation with their customers, to convince them to manage technical debt.

\begin{table*}[ht]
\begin{center}
\includegraphics[width=0.55\textwidth]{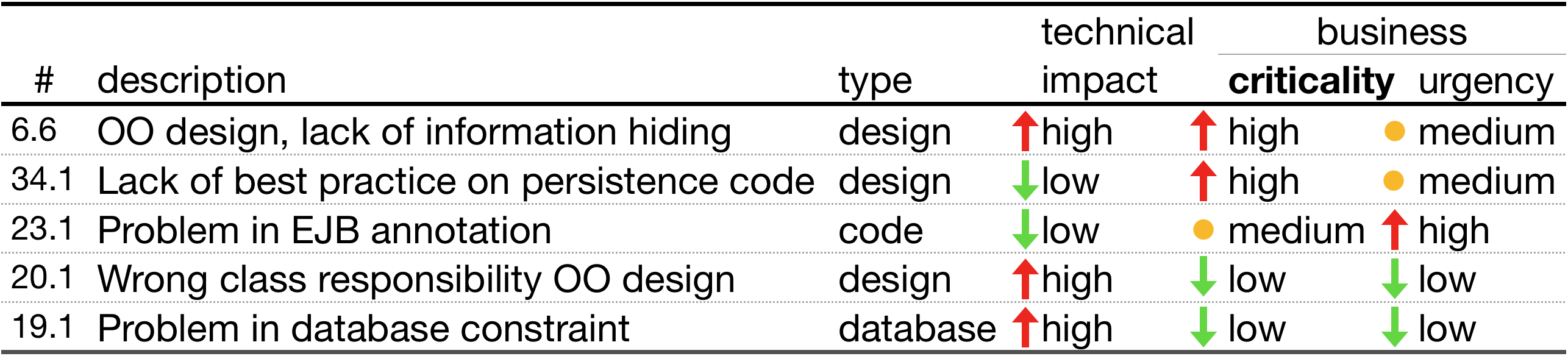}
\caption{Example of technical debt items ordered by their business criticality - Case A}
\label{table_itemExample}
\end{center}
\end{table*}

The cases where we have a low expectation from the business perspective and a high or medium technical priority (75\% for criticality and urgency) may be a source of overestimation of a technical issue. For example, item \#20.1 in Table~\ref{table_itemExample} refers to a low-level design issue, which affects the system maintainability, classified as medium cost and high interest. This item affects a system which supports the ``Request for Payment'' subprocess, with a low criticality and urgency, from the business perspective. As a result, solving this item could be delayed compared to the item \#23.1, with low interest, in technical terms. Item \#23.1, in Table~\ref{table_itemExample}, describes a simple annotation problem, at the code level, which is easy to solve and could have low interest. But, since it affects systems which support the Payment business process, it would have a higher priority.

\subsection{RQ 2. Does the business perspective captured through business process management facilitate the prioritization of technical debt?}

We presented the business process modeling together with the prioritization of technical debt items to two business and two IT stakeholders (Case A). The IT stakeholders declared that the business process visualization was useful to support technical debt prioritization. They also argued that \textit{``many times a critical technical debt must be prioritized even if it affects a low critical business process, to reduce the problem of accumulating debt''}. Indeed, the business prioritization is not a silver bullet to define technical debt prioritization, but it provides an important perspective to help in decision making. 

\begin{table*}[ht]
\begin{center}
\includegraphics[width=0.7\textwidth]{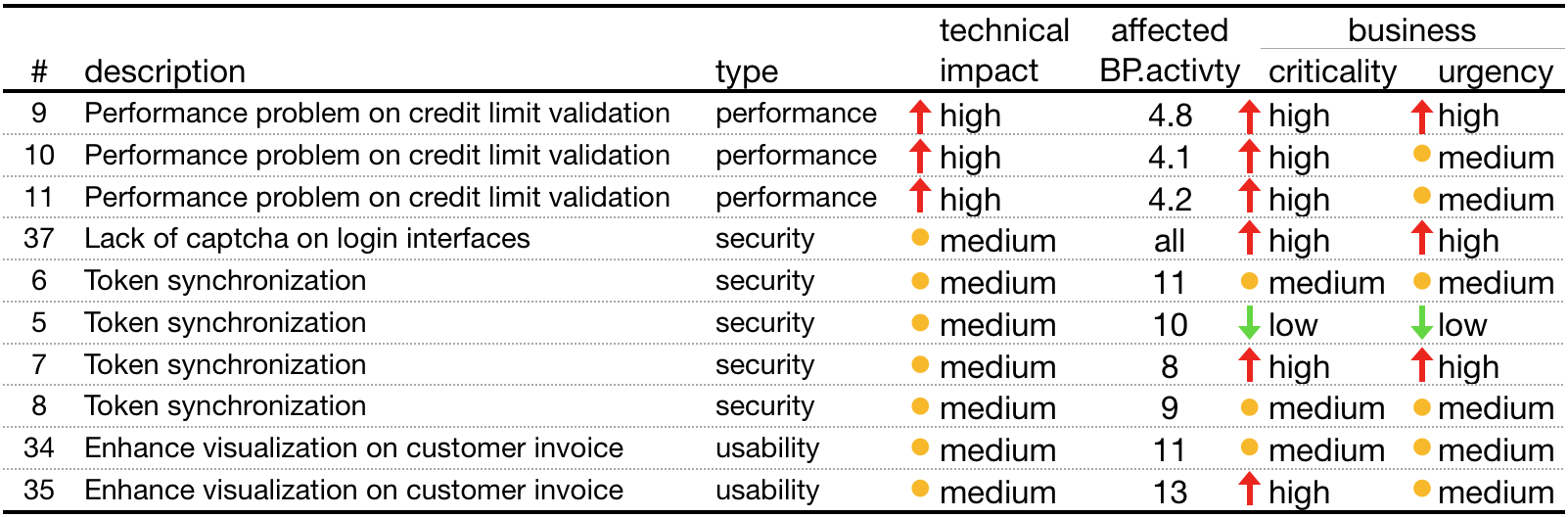}
\caption{Final prioritization after the discussion between business and IT stakeholders - Case B}
\label{table_finalPriority}
\end{center}
\end{table*}

The IT stakeholders also argued that ``\textit{sometimes it is difficult to convince business stakeholders about the risk of acquiring debt, to meet a proposed tight business schedule}''. ``\textit{It is easier to argue with them that it is necessary to solve a low-level design issue} (as described in Table~\ref{table_itemExample} item \#6.6), \textit{since it is explicit that it affects a critical business process}''. With a common language and a proper relationship between business processes and technical debt items to handle technical debt management, the communication between the development team and the business stakeholders can be facilitated.

The final focus group, run in Case B with one business and one technical stakeholder (section \ref{feedback}), showed the influence of the business perspective on decision making. Table \ref{table_finalPriority} shows the resulting prioritization. Four technical debt items which affect all business processes were selected. Both focus group participants used their knowledge about business processes as basis for their argumentation. They tried to convince each other by explaining why a particular technical debt item should be selected. Twice, the business stakeholder had to explain details about the business procedures to convince the technical stakeholder. In the end, all selected technical debt items affected a high or medium business criticality and urgency, and the participants selected high and medium technical impacts. The exception was technical debt item 5, which affects a low priority business process, but participants argued that the effort to replicate the ``Token synchronization'' solution would be low and the gain to pay the debt justified the decision. Another insight was that the business stakeholder used only the information about the business impact and its criticality to make his decisions, i.e., the business perspective captured using business processes was a good basis for decision making.

Finally, in Case A, the account manager explained that using the business perspective for prioritizing technical debt could also provide an objective way to define policies regarding technical debt. Besides the prioritization activity, this can help in decision making about the creation of debt items. Depending on the business process criticality, it would be feasible to deny any high or medium impact debt on it. Her argument leads us to sketch a new approach to prioritize technical debt items based on the business perspective using business process modeling. The next section introduces this approach.

\section{Discussion and Proposed Approach}

This section discusses the findings from the case study which illustrate how aspects of technical debt management and business process modeling can be accomplished in practice.

\subsection{The tension between technical and business perspectives} \label{sec.discussion}

The definition of the technical debt metaphor was consistent among the participants on both cases. The examples they provided showed that they had a solid understanding of the theme. The different roles offered varied perspectives about the sources of technical debt. For example, while the senior developer described issues related to low-level coding and developer behavior, the service manager focused on high-level debt, such as architectural debt and infrastructure debt. All participants were unanimous about the interference of business priorities as a source of technical debt. Tight schedules and service incidents were also identified as sources of technical debt.

The technical leaders and software architects are responsible for estimating the impact of technical debt. They have in-depth technical knowledge and experience and a broad perception of the technologies, dependencies, and requirements. Impact is a perception of the debt's dependencies, integration, technical requirements, the probability of resulting in an error or affecting availability, performance, and so on. In summary, all of these aspects are directly and indirectly related to the question, \textit{``What bad things can happen if I do not pay this debt?''}, a technical leader said. She also said that \textit{``it is difficult to evaluate it objectively. Sometimes a line of code can have high impact, and a whole system or module can have low impact. Sometimes we even have to consider the personality of the colleague responsible for a system we depend on. Technically the solution may be straightforward, but aspects such as organization hierarchy and departmental relationships, for example, can greatly increase the impact of debt''}.

From the business side, the account manager from Case A explained that \textit{``sometimes the level of detail from one side and the lack of proper understanding from the other can influence the decision about how urgent or critical a technical debt item is''}. For example, sometimes the technical team discusses a problem at a low level and presents the problem in terms that the business stakeholders cannot understand. \textit{``Sometimes, when the argumentation is lost, the tech guys still come up with some security trouble to convince everyone''}, the business analyst commented. She also pointed out that \textit{``on the other hand, sometimes business stakeholders use `obscure' motivations to justify a tight schedule or to impose that a feature is more important than solving a technical issue which must wait''}.

A business stakeholder from Case B identified the lack of a broader view of the business as an important problem. She said that \textit{``nowadays people are getting very specialized in their areas and it is difficult to get an overall perspective of the business''}. \textit{``It is quite common to define a scope of a system with one business area, and when we start delivering the releases, some conflicts arise with another business area''}. 

\subsection{Proposed approach}

\begin{figure}[ht]
\begin{center}
\includegraphics[width=0.7\columnwidth]{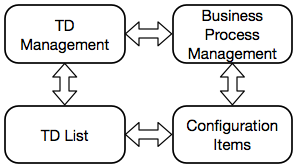}
\caption{Proposed approach.}
\label{fig_approach}
\end{center}
\end{figure}

As a result of answering our two research questions, we outline a preliminary conceptual approach which can contribute to technical debt management using the business perspective. The approach extends existing research work by Guo and Seaman~\cite{Guo:2011, seaman:guo:2011}. Figure~\ref{fig_approach} shows an overview of its components. Now, besides the technical debt management centered on the technical debt list, there are two new areas: business process management and configuration items. The business process management involves the complete lifecycle of the business processes and a set of management tools to deal with strategic, tactical, and operational aspects of the business perspective.

To apply the approach to technical debt prioritization, it is necessary to:
\begin{enumerate}
 \item Keep track of a technical debt list;
 \item Relate the technical debt items to software and/or infrastructure configuration items;
 \item Model the business processes which are supported by the configuration items -- technical artifacts of the systems;
 \item For each business activity: identify which business aspects contribute to decision making (criticality, urgency, financial aspects, etc.);
 \item For each business activity: prioritize the activities considering business objectives;
 \item Conduct the technical debt prioritization, using the business perspective.
\end{enumerate}

\section{Limitations}

While our case studies found evidence that it is possible to align technical debt prioritization with business objectives through business process modeling and mapping of technical debt items to business processes, we cannot generalize our findings to other cases. Other companies might have different characteristics and different processes. Naturally, the number of participants we were able to talk to is also limited. However, we note that twenty-two senior professionals participated in this work, on both cases (twelve on Case A and ten on Case B) and that these individuals played different roles in the company (e.g., senior developers, architects, and business stakeholders). Many of the professionals had previous experience with other companies, giving them a broad perspective on the decisions and opinions.

The companies at which this multiple-case study was conducted employ together more than 1000 developers and build complex systems which affect many people and private companies. The approach used to dealing with technical debt by the participant teams is mature, e.g., both teams use specific tools for the management of technical debt. We are therefore optimistic that these cases can be applied to other teams which have a direct business impact. Validating this will be part of our future work. Indeed, we have received very positive feedback from both companies to continue working together on the evolution of a business-oriented approach to managing technical debt by interacting with other teams and projects. 

Our multiple-case study shows evidence that the information about business priorities can change the way companies make technical debt management decisions. In this paper, we focus on one technical debt management activity: prioritization. However, there are opportunities to explore other technical debt management activities in the future and their interplay with the business perspective, such as identification, measurement, monitoring, and communication~\cite{Rios2018}. There are also opportunities to explore the business process management, by considering other levels of business decision making, such as operational, tactical, and strategic, and other phases of the business process management lifecycle.

\section{Related Work}

Recent research on technical debt has pointed out the lack of a proper business treatment for technical debt management activities~\cite{MARTINI2018, Alves:2016:IMT:2873074.2873279, FERNANDEZSANCHEZ201722}.

Guo and Seaman~\cite{Guo:2011} performed a case study on the release planning of a software application for mobile platforms. They propose a technical debt management framework which considers the principal, the interest amount, and the interest probability when dealing with the technical debt analysis. The scenario consists of an analysis of decision making in the release planning, where a change should be done at a certain point in time and is delayed due to ``time-to-market'' reasons. The work calculates the probability of the interest by asking experts and the implementation effort is measured in staff-hours. The results show that the use of a technical debt management approach could change key decisions in release management and could avoid the negative effects of the debt. Despite the cost related to the software development team, the business value of the two decisions presented in the work could generate more value compared to the cost incurred through the negative impact on the software side.

There are also researchers who consider business metrics and business values to deal with technical debt. Yli-Huumo et al.~\cite{Yli-Huumo2015}, for example, conducted a case study with four companies to understand the relationship between ``Business Model Experimentation'' and technical debt. Business Model Experimentation is a way to perform business model innovation. This approach is based on the Lean model, promoting business model changes in short life cycles. The authors also argue, based on a literature review, that the relationship between technical debt and business models is not well-studied and requires more examination. The authors performed semi-structured interviews with practitioners from four companies and found that those who use business model experimentation reduce intentional technical debt. This finding is an insight into how involving business stakeholders in the process of technical debt management could increase business value.

Our work also considers a business perspective to support decision making on technical debt. We focus on the prioritization activity, and -- unlike related work -- we use business process management to help bridge the gap between the technical perspective and the business perspective. To the best of our knowledge, this is the first work which brings these two disciplines together: technical debt management and business process management.

\section{Conclusions}

This multiple-case study addressed the following research questions: RQ 1. How can the business perspective influence the prioritization of technical debt? RQ 2. Does the business perspective captured through business process management facilitate the prioritization of technical debt?

To address these research questions, we performed a multiple-case study in two large software development companies where we observed how the business perspective can affect the prioritization of technical debt. In particular, we have considered how specific business processes and their respective priorities in terms of business urgency and criticality can change the technically oriented prioritization of debt items. Based on the results of the two cases, we make the following contributions:

\begin{itemize}
\item We found that the business perspective can affect the prioritization of technical debt items;
\item We found that business processes can facilitate the communication and prioritization of technical debt items;
\item We extended the analysis model presented by Rios et al.~\cite{Rios2018} by adding the link between technical debt items and the business processes;
\item We extended the work by Guo and Seaman~\cite{Guo:2011, seaman:guo:2011} to propose a conceptual model to support technical debt management decisions while taking business process management into account.
\end{itemize}

As part of future work, we aim to conduct surveys and interviews with other companies, to add to the results found in this study. We also plan to expand the application of the proposed approach to create and possibly automate a better alignment between technical debt management activities and business expectations. Another interesting area of future work is to investigate the relationship between the granularity of configuration items and technical debt categories as well as the impact of the age of technical debt items and configuration items.

\bibliographystyle{IEEEtran}
\bibliography{localbib.bib}

\end{document}